\documentclass[]{aa}
\usepackage{graphicx}
        
\begin{document}

\title{The gravitational wave radiation of pulsating white dwarfs
       revisited: the case of BPM~37093 and PG~1159-035}

\author{E. Garc\'\i a--Berro$^{1,2}$, 
        P. Lor\'en--Aguilar$^{1,3}$,
        A.  H.  C\'orsico$^{4,5}$\thanks{Member  of   the  Carrera  del 
        Investigador   Cient\'{\i}fico   y   Tecnol\'ogico,    CONICET, 
        Argentina.}, 
        L. G. Althaus$^{1\star}$,\\
        J. A. Lobo$^{2,3}$, and
        J. Isern$^{2,3}$}

\offprints{E. Garc\'\i a--Berro}

\institute{$^1$Departament  de    F\'\i  sica   Aplicada,  Universitat
           Polit\`ecnica  de Catalunya,  Av.  del  Canal  Ol\'\i mpic,
           s/n, 08860, Castelldefels, Barcelona, Spain\\
           $^2$Institut d'Estudis  Espacials de Catalunya,  Ed. Nexus,
           c/Gran Capit\`a 2, 08034, Barcelona, Spain\\
           $^3$Institut  de  Ci\`encies de  l'Espai, C.S.I.C.,  Campus 
           UAB, Facultat de  Ci\`encies, Torre  C-5, 08193 Bellaterra,
           Spain\\
           $^4$Facultad de  Ciencias Astron\'omicas y Geof\'{\i}sicas,
           Universidad  Nacional de  La Plata,  Paseo del  Bosque s/n,
           (B1900FWA) La Plata, Argentina\\
           $^5$Instituto  de   Astrof\'{\i}sica   La  Plata,     IALP, 
           CONICET-UNLP\\
\email{garcia@fa.upc.edu,                             loren@fa.upc.edu,
acorsico@fcaglp.unlp.edu.ar,    leandro@fa.upc.edu,   lobo@ieec.fcr.es,
isern@ieec.fcr.es} }

\date{Received; accepted}

\abstract{We  compute  the emission  of  gravitational radiation  from
pulsating white dwarfs.   This is done by using  an up-to-date stellar
evolutionary  code coupled with  a state-of-the-art  pulsational code.
The emission of gravitational waves  is computed for a standard $0.6\,
M_{\sun}$  white  dwarf  with   a  liquid  carbon-oxygen  core  and  a
hydrogen-rich envelope, for a massive  DA white dwarf with a partially
crystallized core for which  various $\ell=2$ modes have been observed
(BPM~37093) and for PG~1159-035, the  prototype of the GW~Vir class of
variable stars, for which  several quadrupole modes have been observed
as well.  We find that these  stars do not radiate sizeable amounts of
gravitational  waves through  their observed  pulsation  $g$-modes, in
line  with  previous studies.   We  also  explore  the possibility  of
detecting  gravitational  waves  radiated  by  the  $f$-mode  and  the
$p$-modes.  We find that in this case the gravitational wave signal is
very large and, hence, the  modes decay very rapidly.  We also discuss
the  possible implications of  our calculations  for the  detection of
gravitational waves  from pulsating white dwarfs  within the framework
of future space-borne interferometers like LISA.
 }

\authorrunning{Garc\'\i a--Berro et al.}

\titlerunning{Gravitational wave radiation of pulsating white
dwarfs}

\maketitle


\section{Introduction}

Gravitational   waves   are    a   direct   consequence   of   General
Relativity. Many efforts have been done so far to detect them, but due
to the intrinsic experimental  difficulties involved in the process of
detection and  in the  data analysis no  definite result has  yet been
obtained.  Supernova  core collapse, binary  systems involving compact
objects, and  pulsating neutron  stars are, amongst  others, promising
sources  of   gravitational  waves  ---   see  Schutz  (1999)   for  a
comprehensive review on the subject.  Moreover, with the advent of the
current generation  of terrestrial gravitational  wave detectors, like
LIGO (Abramovici et al.  1992),  VIRGO (Acernese et al.  2004), GEO600
(Willke  et al.   2004),  or TAMA  (Takahashi  et al.   2004), and  of
space-borne  interferometers like  LISA (Bender  et al.   1998, 2000),
gravitational wave astronomy will probably be soon a tangible reality.

Despite of its potential interest, the emission of gravitational waves
by  pulsating white dwarfs  has been  little explored  up to  now.  In
fact, apart  from the pionering work  of Osaki \&  Hansen (1973), only
the gravitational  wave radiation of rotating  white dwarfs undergoing
quasi-radial oscillations has been  studied so far --- see Benacquista
et al. (2003) and references therein. White dwarfs are the most common
end-point  of  the  evolution  of low-  and  intermediate-mass  stars.
Hence,  white dwarfs  constitute, by  far, the  most  numerous stellar
remnants  in our  Galaxy,  outnumbering neutron  stars. Moreover,  the
relative simplicity of their physics allows us to obtain very detailed
models  which   can  be   ultimately  compared  with   their  observed
properties.  Among  white dwarfs there are three  specific families of
variable stars, known as ZZ~Ceti (or DAV, with hydrogen-rich envelopes
and $T_{\rm eff} \sim  12\,000$~K), V777~Her (or DBV, with helium-rich
envelopes  and $T_{\rm  eff}  \sim 25\,000$~K)  and  GW~Vir stars  (or
variable  PG~1159 objects, with  envelopes which  are rich  in carbon,
oxygen and  helium, and $T_{\rm  eff}$ ranging from $\sim  80\,000$ to
$150\,000$~K), which  show periodic  variations in their  light curves
--- see  Gautschy  \& Saio  (1995;  1996)  for  reviews.  The  typical
periods are within $\sim 100$~s and $\sim 2\,000$~s and, consequently,
lay in the region of frequencies to which LISA will be sensitive.  The
luminosity  changes of  these  variable stars  have been  successfully
explained as due to  nonradial $g$-mode pulsations.  At present, there
is a general consensus that variable white dwarfs are very interesting
targets   for  pulsational  studies.    Their  very   simple  internal
structures   allow  us  to   predict  theoretically   the  pulsational
frequencies  with a  very high  degree of  detail  and sophistication.
Also, they have a very rich  spectrum of frequencies which may give us
information about the stellar mass,  the core composition, the mass of
the surface helium and hydrogen layers (if present), the angular speed
of  rotation and  the  strength of  the  magnetic field  --- see,  for
instance, Pfeiffer  et al.  (1996)  and Bradley (1998,  2001), amongst
others.  Consequently,  it is  not surprising that  in the  last years
ZZ~Ceti and V777~Her white dwarfs,  as well as GW~Vir stars, have been
the preferred targets for the network called ``Whole Earth Telescope''
(WET).  WET observations  have been of an unprecedent  quality, and in
some cases have  allowed us to disentangle the  internal structure and
evolutionary  status of  several  white dwarf  stars  by applying  the
powerful tools of asteroseismology (Nather 1995; Kawaler 1998).

BPM~37093 is the most massive pulsating white dwarf ever found (Kanaan
et  al.  1992).  It  is a  massive ZZ~Ceti  star ---  that is,  with a
hydrogen-rich  atmosphere ---  with a  stellar mass  of $\sim  1.05 \,
M_{\sun}$, and an effective temperature $T_{\rm eff}\simeq 11\,800$~K.
BPM~37093  has   been  thoroughly  studied   (both  theoretically  and
observationally)  because   presumably  it  should   have  a  sizeable
crystallized core (Winget et al.  1997).  Hence, for BPM~37093 we have
detailed models (Montgomery \& Winget 1999; C\'orsico et al. 2005) and
extensive  observational  data (Kanaan  et  al. 2005).   Interestingly
enough,  one of  the most  apparent modes  of BPM~37093  has  a period
$P=531.1$~s, very  close to frequency  of maximum sensitivity  of LISA
and pulsates  with $\ell=2$.   It should be  noted at this  point that
$\ell=1$ modes  do not radiate  gravitational waves and  that $\ell=2$
modes  are relevant  for  the emission  of  gravitational waves,  thus
making BPM~37093 a especially suitable target for LISA.  Moreover, the
distance  to  BPM~37093   is  known  ($d=16.8$~pc).   Consequently,  a
detailed study of the  possibilty of detecting the gravitational waves
emitted by this star is of  the maximum interest, but still remains to
be done.  On the other  hand, PG~1159-035, the prototype of the GW~Vir
class of objects,  has a complex spectrum with  several $\ell=2$ modes
(Winget  et  al.  1991).   Unfortunately  there  is  not any  reliable
parallax  determination for  PG~1159-035.  More  specifically, Werner,
Heber \&  Hunger (1991) provide  $d\sim 800^{+600}_{-400}$~pc, whereas
Kawaler \& Bradley (1994)  obtained $d\simeq 400\pm 40$~pc. However, a
spectroscopic  determination  of  its  mass  ($M_\star  \simeq  0.54\,
M_{\sun}$) is available.  These are, to the best of our knowledge, the
only  two  known  white  dwarf  pulsators  with  confirmed  quadrupole
$g$-modes.

In  this paper  we compute  the emission  of gravitational  waves from
pulsating  white  dwarfs. We  first  compute  the gravitational  waves
radiated   by  a  typical   $0.6\,  M_{\sun}$   white  dwarf   with  a
carbon-oxygen core and a  $10^{-4}\, M_\star$ hydrogen envelope, which
we regard as  our fiducial model. For this model  white dwarf we first
compute the gravitational waves  emitted by $g$-modes. Then we compute
the gravitational waves emitted by BPM~37093 and PG~1159-035, the only
two known white dwarfs with  quadrupole $g$-modes. As it will be shown
below, we find that the fluxes radiated away by these two stars in the
form of gravitational waves are very  small. This is the reason why we
also explore  other possibilities. In  particular we also  compute the
fluxes radiated by  the $f$- and $p$-modes, independently  of the lack
of  observational   evidence  for  these  modes   in  pulsating  white
dwarfs. The paper is organized as  follows.  In \S 2 we briefly review
the basic  characteristics of nonradial  pulsation modes.  In \S  3 we
discuss the  numerical codes used  to compute the  nonradial pulsation
modes of the white dwarf models presented here. Sect.  4 is devoted to
obtain the  expressions for the  emission of gravitational  waves from
pulsating white dwarfs. Finally in \S 5 we present our results whereas
in \S 6 we summarize our findings and we draw our conclusions.

\section{Nonradial pulsation modes}

While  we  highly  recommend  the  interested reader  to  consult  the
excellent monographs of Unno et  al. (1989) and Cox (1980) for details
of nonradial stellar  pulsations, we believe that a  brief overview of
the  basic  properties of  nonradial  modes  it  is worthwhile  to  be
done. Briefly,  nonradial modes are  the most general kind  of stellar
oscillations.   There exist  two subclasses  of  nonradial pulsations,
namely, {\sl spheroidal} and {\sl toroidal} modes. Of interest in this
work are the spheroidal modes, which are further classified into $g$-,
$f$- and $p$-modes according to the main restoring force acting on the
oscillations, being  gravity for the  $g$- and $f$-modes  and pressure
for the $p$-modes.

For a  spherically symmetric star,  a linear nonradial  pulsation mode
can   be    represented   as   a    standing   wave   of    the   form
$\Psi^\prime_{k,\ell,m}(r,\theta,\phi,t)= \Psi^{\prime}_{k,\ell, m}(r)
\; Y^m_{\ell}(\theta,\phi)\;  {\rm e}^{i \sigma_{k,\ell,m}  t}$, where
the symbol `` $^\prime$ ''  indicates a small Eulerian perturbation of
a given  quantity $\Psi$ (like the  pressure, gravitational potential,
etc)  and $Y^m_{\ell}(\theta,\phi)$  are  the corresponding  spherical
harmonics.  Geometrically, $\ell$ is the  number of nodal lines in the
stellar  surface  and  $m$  is  the  number of  such  nodal  lines  in
longitude.  In absence of any  physical agent able to remove spherical
symmetry  (like  magnetic fields  or  rotation), the  eigenfrequencies
$\sigma_{k,\ell,m}$ are  dependent upon $\ell$ but are $2\ell+1$ times
degenerate in $m$.  Finally, $\Psi^\prime_{k,\ell,m}(r)$ is the radial
part  of the  eigenfunctions, which  for realistic  models necessarily
must be  computed numerically together  with $\sigma_{k,\ell,m}$.  The
index $k$ (known  as the radial order of the  mode) represents, in the
frame of simple stellar models  (like those of white dwarf stars which
we shall study below), the number  of nodes in the radial component of
the eigenfunction. Generally  speaking, $g$-modes are characterized by
low oscillation frequencies (long periods) and by displacements of the
stellar fluid  essentially in the horizontal  direction.  At variance,
$p$-modes have high frequencies  (short periods) and are characterized
by essentially  radial displacements  of the stellar  fluid.  Finally,
there is a single $f$-mode for a given $\ell$ ($ \geq 2$) value.  This
mode  does not  have any  node in  the radial  direction ($k=  0$) and
possesses an  intermediate character  between $g$- and  $p$-modes.  In
fact, its eigenfrequency  lies between that of the  low order $g$- and
$p$-modes, and generally slowly  increases when $\ell$ increases.  For
$g$-modes  ($p$-modes), the  larger  the  value of  $k$  is the  lower
(higher) the oscillation frequency.
  
\section{Numerical codes} 

We compute  the nonradial  pulsation modes of  the white  dwarf models
considered in  this work  with the help  of the same  pulsational code
described in detail in C\'orsico at al. (2001a, 2002). The code, which
is  based  on a  standard  finite  differences  scheme, provides  very
accurate oscillation frequencies and nonradial eigenfunctions, and has
been employed in numerous works on white dwarf pulsations --- see, for
instance, C\'orsico  et al.  (2004) and references  therein.  The code
solves the fourth-order set  of equations governing Newtonian, linear,
nonradial stellar pulsations  in the adiabatic approximation following
the dimensionless formulation given in  Unno et al.  (1989).  To build
up the white dwarf models  needed for our pulsational code we employed
the LPCODE  evolutionary code  described in detail  in Althaus  et al.
(2003, 2005).   Our evolutionary code contains  very detailed physical
ingredients.  A full description  of these physical ingredients is out
of the scope  of this paper and, consequenlty,  the reader is referred
to Althaus et al.  (2003,  2005) for an extensive description of them.
Instead, we will  only summarize here the most  important inputs.  For
instance, the equation of state includes partial ionization, radiation
pressure,  ionic  contributions,  partially degenerate  electrons  and
Coulomb  interactions.  For  the  white dwarf  regime,  we include  an
updated  version of  the  equation  of state  of  Magni \&  Mazzitelli
(1979).  The code uses  OPAL radiative opacities --- including carbon-
and  oxygen-rich  compositions  ---  for  arbitrary  metallicity  from
Iglesias \&  Rogers (1996) and  molecular opacities from  Alexander \&
Ferguson  (1994).  High-density  conductive opacities  are  taken from
Itoh et  al. (1994) and  the references cited there,  whereas neutrino
emission  rates  are those  of  Itoh  et  al. (1996),  and  references
therein. It  is important  to mention at  this point that  the stellar
models for BPM~37093 and PG~1159-035 discussed below have been derived
from full evolutionary calculations that take into account the history
of  the progenitor  stars ---  see Althaus  et al.   (2003,  2005) for
details.  Finally,  it is worth noting  as well that  during the white
dwarf cooling  phase, the effects of  time-dependent element diffusion
have been considered in the calculations.

\section{Gravitational waves from a pulsating white dwarf}

Although  the  basic formalism  for  deriving  the gravitational  wave
radiation of pulsating objects  (either white dwarfs or neutron stars)
is well  known --- see,  for instance, Osaki  \& Hansen (1973)  --- we
consider   appropriate  to  summarize   it  here   for  the   sake  of
completeness.  Moreover we will extend it to the case in which a white
dwarf  has a  partially  crystallized core.   Generally speaking,  the
amplitude of a gravitational  wave emitted from any slow-moving source
in the quadrupole approximation is given by (Misner et al. 1973)

\begin{equation} 
h^{\rm TT}_{ij}=\frac{2G}{c^4d} \ddot{Q}^{\rm TT}_{ij},
\end{equation}

\noindent where ``TT'' stands  for the traceless-transverse gauge, $d$
for  the  distance, and  $Q$  is the  quadrupole  moment  of the  mass
distribution, which is defined as

\begin{equation} Q_{ij}=\int_{R^3} \rho (\vec{r})
(3x_ix_j-\delta_{ij}r^2)d^{3}r
\end{equation}

\noindent  As  previously  stated,  we  assume that  the  spatial  and
temporal behavior of the the  perturbed density profile is provided by
the following expression:

\begin{equation} \rho(\vec{r},t)=\rho_0(r) + \rho^\prime(r)
{\rm Re}\left( Y^m_{\ell}(\theta,\phi) {\rm e}^{i \sigma t}\right)
\end{equation}

\noindent   where  $\rho_0$  is   the  unperturbed   density  profile,
$\rho^\prime(r)$  stands for  the radial  perturbation of  the density
profile,  $Y^m_{\ell}(\theta,\phi)$ are  the spherical  harmonics, and
$\sigma$ is the pulsational frequency.  As we are dealing with $\ell =
2$ modes, and  since the emission of gravitational  waves in this case
will  be the  same for  all the  values of  $m$, we  shall  choose the
simplest case. That  is, we adopt $\ell=2$ and  $m= 0$.  Additionally,
it  must  be  taken  into   account  that  BPM~37093  has  a  sizeable
crystallized  core.  Therefore,  the appropriate  boundary  conditions
differ  from those of  an ordinary  star.  Particularly,  the boundary
condition at the stellar center  (when crystallization has not yet set
in) is that  given by Osaki \& Hansen (1973).   However, when the core
of  the white  dwarf  undergoes crystallization  we  switch the  fluid
internal boundary conditions to the so-called ``hard sphere'' boundary
conditions (Montgomery \& Winget 1999).  Within this approximation the
nonradial   eigenfunctions   are  inhibited   to   propagate  in   the
crystallized  region of the  core. Consequently,  and keeping  in mind
that for the axisymmetric case $Q_{11}=Q_{22}=-\frac{1}{2} Q_{33}$ and
$Q_{ij}=0$  if  $i\neq  j$  (Osaki  \&  Hansen  1973),  the  following
expression can be easily obtained

\begin{figure*}[t]
\centering
\includegraphics[clip,width=360pt]{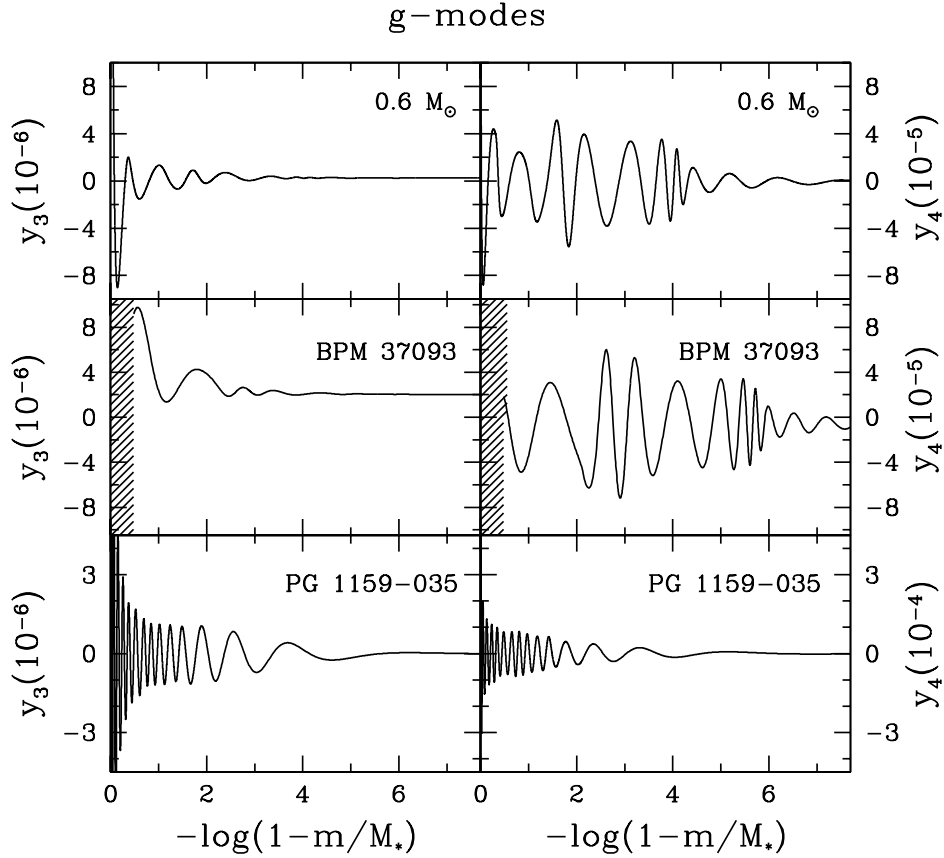}
\caption{Run  of the $y_3$  --- left  panels ---  and $y_4$  --- right
panels  ---  eigenfunctions for  the  $g$-modes  of  a typical  $0.6\,
M_{\sun}$ white  dwarf ---  top panels ---  for BPM~37093  --- central
panels --- and for PG~1159-035 ---  bottom panels --- as a function of
the mass  coordinate $\log (1-m/M_\star)$.  For the  case of BPM~37093
the  crystallized  core is  shown  as a  hatched  area.  See text  for
details.}
\label{fig1} 
\end{figure*}

\begin{figure*} 
\centering
\includegraphics[clip,width=360pt]{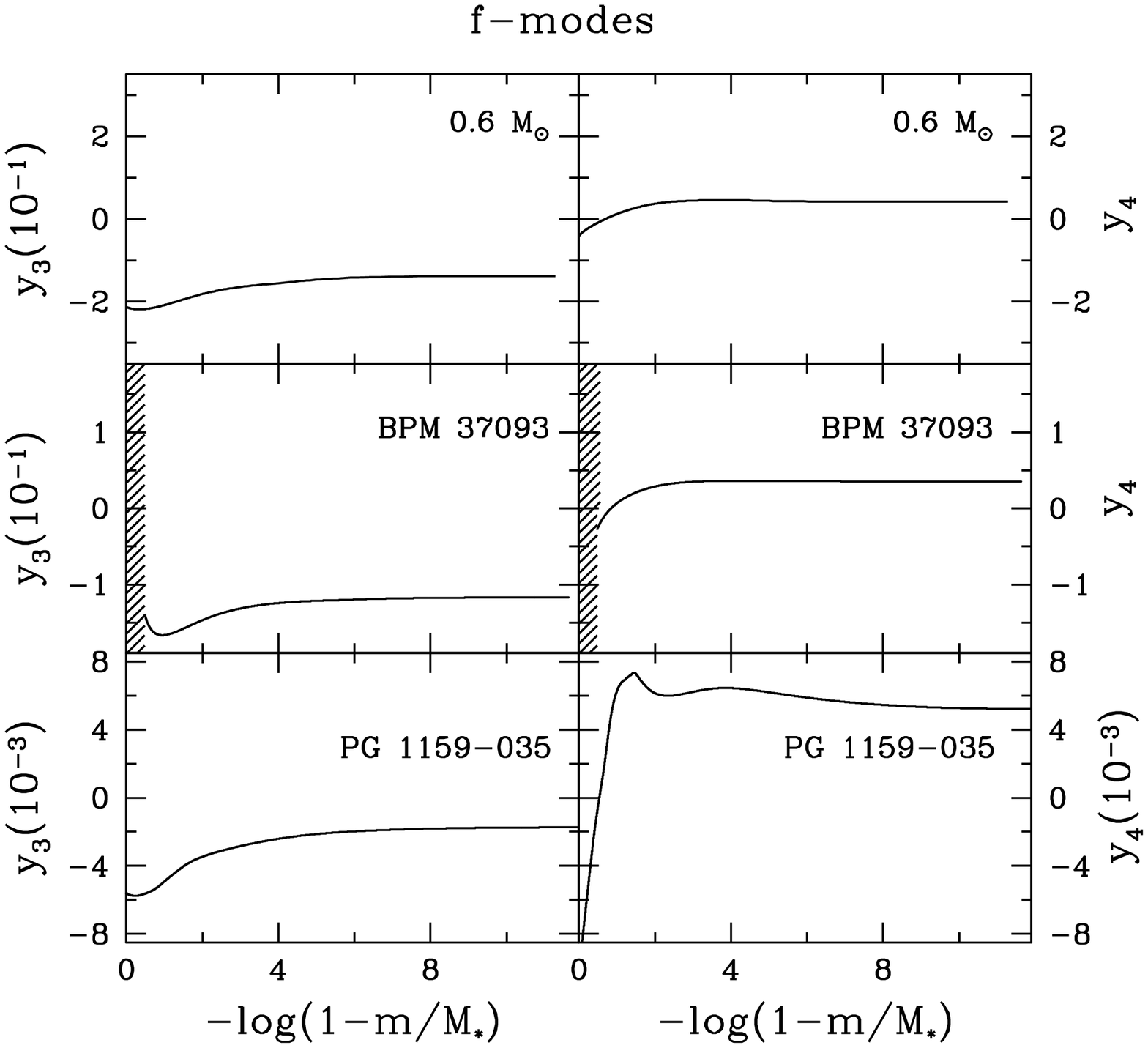}
\caption{Same  as  Fig.~1 for  the  $f$-mode  of  our $0.6\, M_{\sun}$
fiducial model, for BPM~37093 and for PG~1159-035.}
\label{fig2}
\end{figure*}

\begin{eqnarray} 
Q_{33}=&\,&\int^{2 \pi}_0 d \phi \int^{\pi}_0 
\sqrt{\frac{5}{16\pi}}
\sin\theta (\cos^2\theta -1)
(3\cos^2\theta-1)d\theta\nonumber\\
&&\int^{R_\star}_{R_0}r^4\rho^\prime(r)\cos(\sigma t) \, dr
\end{eqnarray} 

\noindent where $R_{\star}$  and $R_0$ are the stellar  radius and the
radial  coordinate  of the  crystallization  front, respectively.   By
using Poisson's equation

\begin{equation} 4\pi
G\rho^\prime(r)=\frac{1}{r^2}\frac{d}{dr}\Big(r^2\frac{d{\Phi^\prime}}{dr}
\Big)- \frac{6}{r^2}\Phi^\prime
\end{equation}
 
\begin{figure*} 
\centering
\includegraphics[clip,width=360pt]{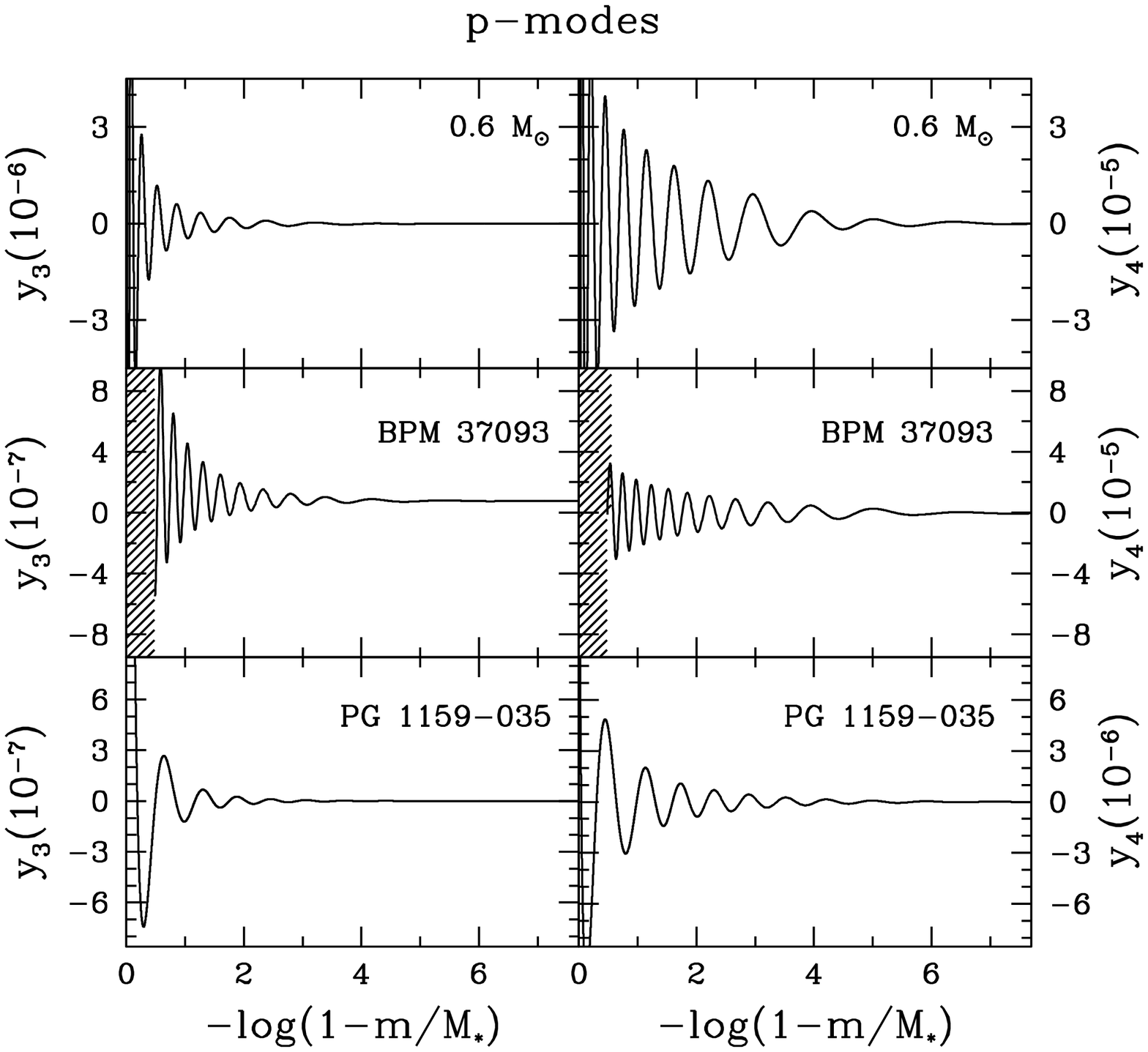}
\caption{Same  as  Fig.~1 for  the  $p$-modes of  our $0.6\, M_{\sun}$
fiducial model, for BPM~37093 and for PG~1159-035}
\label{fig3}
\end{figure*}

\noindent  and  Eq.~(4)  it  can  be  shown  after  a  straightforward
calculation that the dimensionless strain, $h^{\rm TT}_{33}$, is given
by

\begin{eqnarray} 
h^{\rm TT}_{33} \approx &\,& 5\times 10^{-18}
\Big(\frac{M_\star}{M_{\sun}}\Big)
\Big(\frac{R_\star}{R_{\sun}}\Big)^2 
\Big(\frac{\nu}{1\,{\rm mHz}}\Big)^2
\Big(\frac{1 \, {\rm pc}}{d}\Big)\nonumber\\
&&\Big[A(R_\star)-F_\mu F_R^2 A(R_0)\Big]\cos(\sigma t)
\end{eqnarray}

\noindent being $\nu=2\pi\sigma$ the frequency of the signal and

\begin{eqnarray} 
A(r) &\equiv& y_4-2y_3 \nonumber\\
y_3 &\equiv& \frac{\Phi^\prime}{gr}  \nonumber\\
y_4 &\equiv& \frac{1}{g} \frac{d\Phi^\prime}{dr}\\
F_\mu &\equiv& \frac{M_0}{M_\star} \nonumber\\
F_R &\equiv& \frac{R_0}{R_\star} \nonumber
\end{eqnarray}

\noindent  In   these  expressions  $\Phi^\prime$   is  the  perturbed
gravitational potential,  $g$ is  the gravitational acceleration  at a
given radius, $M_{\star}$  and $R_{\star}$ are the mass  and radius of
the  star, and  $M_{0}$ and  $R_{0}$ are  the mass  and radius  of the
crystallized   core   of  the   white   dwarf.    Obviously,  in   the
non-crystallized  case,  $M_{0}$  and  $R_{0}$  are  identically  zero
(corresponding  to   the  stellar  center).   The   advantage  of  the
previously described formalism is  that the quantities $y_3$ and $y_4$
can be easily tabulated for  a typical stellar model, whereas the rest
can be  observationally obtained. Finally, the  luminosity radiated in
the form of gravitational waves is (Osaki \& Hansen 1973):

\begin{eqnarray}
L_{\rm GW}\approx &\,&10^{36}
\Big(\frac{M_\star}{M_{\sun}}\Big)^2
\Big(\frac{R_\star}{R_{\sun}}\Big)^4
\Big(\frac{\nu}{1\,{\rm mHz}}\Big)^6\nonumber\\
&&\Big[A(R_\star)-F_\mu F_R^2 A(R_0)\Big]^2
\end{eqnarray}

\section{Results}

Figure 1 shows  the run, as a function of the  mass coordinate, of the
functions $y_3$  --- left  panels --- and  $y_4$ --- right  panels ---
discussed in \S  4 for the $g$-modes of our  fiducial model (a typical
$0.6\, M_{\sun}$ white  dwarf made of carbon and  oxygen with a liquid
core  and hydrogen  envelope of  $10^{-4} M_*$),  for BPM~37093  --- a
massive ($M_\star\simeq 1.05\, M_{\sun}$) white dwarf, with a sizeable
crystallized core --- and for PG~1159-035, the other known white dwarf
with unambiguously  identified quadrupole $g$-modes.   It is important
to mention at  this point that the fractional change  in radius due to
pulsations, $\delta R_\star/R_\star$, must not necessarily be the same
for  each pulsation  mode.  However,  the lineal  theory  of nonradial
pulsations does not provide any clue about the value of the fractional
change in  radius, since the  governing equations are  homogeneous and
the normalization  of the eigenfunctions is arbitrary  (Cox 1980).  In
addition,  $\delta  R_\star/R_\star$  is  poorly  constrained  by  the
observations,  since  the  luminosity  variations of  pulsating  white
dwarfs are almost exclusively caused by changes in temperature, not by
their  radius  variations (Robinson  et  al.   1982).   Thus, we  have
adopted, somehow arbitrarily, that the fractional change in radius due
to  pulsations   is  $\delta  R_\star/R_\star=10^{-4}$   for  all  the
considered modes, which is a typical value for pulsating white dwarfs,
and reproduces  reasonably the amplitude  of the observed  light curve
(Robinson et al. 1982).

We start discussing  the results obtained for our  fiducial model (top
panels  of  Fig.~1).   We   have  chosen  to  display  the  quadrupole
($\ell=2$)  $g$-mode with  radial  order $k=25$,  which  has a  period
$P=678.22$~s.  As can be seen  the functions $y_3$ and $y_4$ are small
everywhere in the  star, being their amplitudes of  the order of $\sim
10^{-6}$  and $\sim 10^{-4}$,  respectively. Moreover,  the amplitudes
are only significant for the  central regions of the white dwarf.  For
the case  of BPM~37093 --- central  panels --- we  show the quadrupole
$g$-mode with $k=27$. This $g$-mode has a period $P=536.4$~s, which is
very close to one of  the observed periods, which has $P=531.1$~s.  It
is important to realize that since BPM~37093 is a massive white dwarf,
a sizeable region of its core is crystallized.  This region is clearly
marked in the central panels of  Fig.~1 as a shaded area. We note that
in  this region  the  amplitudes of  both  $y_3$ and  $y_4$ are  null.
Finally, for  the PG~1159-035 model  --- bottom panels ---  we display
the quadrupole mode  with $k=30$, which best fits  the observed period
of $P=423.2$~s.  This mode has a period $P=423.8$~s, thus providing an
excellent fit to the observational data.

\begin{table*}
\centering
\caption{Summary of  the gravitational wave emission  of the $g$-modes
of BPM~37093  and PG~1159-035.  Our  fiducial model is also  shown for
the  sake  of  comparison.   In  all cases  we  have  adopted  $\delta
R_\star/R_\star=10^{-4}$.}
\begin{tabular}{lccccccc}
\hline
\hline
Model    &  $M/M_{\sun}$  &  $k$ & $P_{\rm o}$~(s) & $P_{\rm c}$~(s)  & $h_{\rm max}$ &
      $L_{\rm GW}$~(erg/s) & $\log(E_K)$~(erg)\\
\hline
  BPM~37093        &  1.10  &   26   &  511.7  &  516.7  & $4.8\times 10^{-28}$ &  $9.2\times 10^{18}$  &  46.4\\
                   &        &   27   &  531.1  &  536.4  & $5.4\times 10^{-28}$ &  $1.1\times 10^{19}$  &  46.5\\
                   &        &   28   &  548.4  &  555.8  & $5.3\times 10^{-28}$ &  $9.9\times 10^{18}$  &  46.6\\
                   &        &   29   &  582.0  &  574.9  & $6.4\times 10^{-28}$ &  $1.3\times 10^{19}$  &  46.7\\
                   &        &   30   &  600.7  &  593.0  & $6.4\times 10^{-28}$ &  $1.2\times 10^{19}$  &  46.8\\
                   &        &   32   &  633.5  &  630.4  & $5.5\times 10^{-28}$ &  $8.3\times 10^{18}$  &  46.9\\
                                                   \hline
  PG~1159-035      &  0.54  &   25   &  352.7  &  358.9  & $1.0\times 10^{-30}$ &  $5.0\times 10^{16}$  &  44.0\\
                   &        &   30   &  423.8  &  423.2  & $2.3\times 10^{-30}$ &  $1.7\times 10^{17}$  &  43.8\\
                   &        &   50   &  694.9  &  684.5  & $3.5\times 10^{-31}$ &  $1.6\times 10^{15}$  &  43.4\\
                   &        &   55   &  734.2  &  752.9  & $2.0\times 10^{-31}$ &  $4.1\times 10^{14}$  &  43.3\\
                   &        &   60   &  812.6  &  818.1  & $1.8\times 10^{-31}$ &  $2.8\times 10^{14}$  &  43.2\\
                   &        &   70   &  968.7  &  950.1  & $6.0\times 10^{-32}$ &  $2.5\times 10^{13}$  &  42.8\\
                                                   \hline
 0.6 $M_{\sun}$    &  0.6   &   1    &  ---  &     66.6  & $6.9\times 10^{-25}$ &  $1.0\times 10^{28}$  &  47.0\\
                   &        &  10    &  ---  &  310.3   &  $1.8\times 10^{-27}$ &  $3.1\times 10^{21}$  &  44.8\\
                   &        &  20    &  ---  &  555.2   &  $1.5\times 10^{-27}$ &  $6.6\times 10^{20}$  &  45.1\\

\hline
\hline
\end{tabular}
\end{table*}

In  Table~1  we  summarize  the  most important  results  for  several
$g$-modes  of the models  computed so  far. The  first column  of this
table lists the model. The second column corresponds to its respective
mass.   In the  third column  we show  the radial  order, $k$,  of the
computed $g$-mode. The observed  and the computed periods (in seconds)
are given in columns 4 and  5, respectively.  We have assumed that all
of the observed periods in BPM~37093 and in PG~1159-035 in Table~1 are
$\ell= 2$, following the works by  Kannan et al.  (2005) and Winget et
al.   (1991),   respectively.   In  column  6  we   give  the  maximum
dimensionless  strain,  $h_{\rm max}$,  as  computed  from Eqs.~6  and
7. For BPM~37093 we have computed it adopting the measured distance to
the  source ($d=16.8$~pc),  whereas  for our  fiducial  model we  have
adopted a  distance $d=50$~pc, which we consider  to be representative
of a typical white dwarf.  For the case of PG~1159-035 we have adopted
a distance of 400~pc, in line with the determinations of Werner et al.
(1991) and  Kawaler   \&  Bradley  (1994).   Column   8  provides  the
luminosity in  the form of gravitational waves  radiated away, $L_{\rm
GW}$, computed with  Eq.~(8).  Finally, in the last  column of Table~1
we list  the kinetic energy of each  one of the modes.   Note that, in
general, the  agreement between the computed and  the observed periods
is rather good  for all the modes, both for the  case of BPM~37093 and
for PG~1159-035.  However, the amplitudes of the dimensionless strains
are rather  small in  all the cases.   This is  also the case  for the
luminosities radiated away in the form of gravitational waves.  In the
best of the cases BPM~37093  radiates away $\sim 10^{19}$~erg/s in the
form of gravitational waves,  whereas PG~1159-035 radiates away a much
more modest amount, only $\sim 10^{17}$~erg/s.

It  is interesting  to  note that  for  our fiducial  model ---  third
section of  Table~1 --- the larger  the radial order  $k$, the smaller
the  dimensionless strain  is  and the  smaller  are the  luminosities
radiated away  in the form  of gravitational waves. In  particular, an
increase  from $k=1$ to  $k=10$ produces  a reduction  of a  factor of
almost  $4\times 10^2$  in the  dimensionless strain  and  of $3\times
10^6$  in  the  flux  of  gravitational waves.   The  reductions  when
considering the $k=20$ mode are  much more modest. The reason for this
is the  following, low-$k$  modes sample the  core more  than high-$k$
modes, and  since the core has  a higher density,  larger mass motions
are   produced,  and   hence  more   gravitational  wave   losses  are
produced. Nevertheless, both the  dimensionless strains and the fluxes
of gravitational waves  are in this case much  larger than those found
for BPM~37093 and PG~1159-035.  The reasons for this behavior are easy
to understand.  For  the case of BPM~37093 pulsations  occur only in a
small region of  the star as a result of  its crystallized core. Thus,
despite its mass being much larger than that of our fiducial model the
emission of  gravitational waves is strongly inhibited.   For the case
of  PG~1159-035, the most  important reason  why so  few gravitational
waves are radiated away is its small mass (and average density).

\begin{table*}
\centering
\caption{Summary of  the gravitational wave  emission of the  $f$- and
$p$-modes  of BPM~37093 and  PG~1159-035. Our  fiducial model  is also
shown  for the  sake  of comparison.   In  all cases  we have  adopted
$\delta R_\star/R_\star=10^{-4}$.}
\begin{tabular}{lccccccc}
\hline
\hline
  Model            &  $M/M_{\sun}$  &  Mode & $k$ & $\nu$~(Hz) & $h_{\rm max}$ & $L_{\rm GW}$~(erg/s) & $\log(E_K)$~(erg)\\
\hline
  BPM~37093        &  1.10 &   $f$   &   0   &  $2.7 \times 10^{-1}$   &  $6.6 \times 10^{-19}$   &  $3.4 \times 10^{41}$ & 49.2\\
                   &       &   $p$   &   1   &  $9.3 \times 10^{-1}$   &  $9.1 \times 10^{-20}$   &  $7.7 \times 10^{40}$ & 47.8\\
\hline
  PG~1159-035      &  0.54 &   $f$   &   0   &  $4.8 \times 10^{-2}$   &  $1.2 \times 10^{-22}$   &  $2.1 \times 10^{35}$ & 45.5\\
                   &       &   $p$   &   1   &  $5.9 \times 10^{-2}$   &  $7.1 \times 10^{-23}$   &  $5.1 \times 10^{34}$ & 44.6\\
\hline
  $0.6\, M_{\sun}$ &  0.60 &   $f$   &   0   &  $8.8 \times 10^{-2}$   &  $5.6 \times 10^{-20}$   &  $2.3 \times 10^{39}$ & 47.0\\
                   &       &   $p$   &   1   &  $1.8 \times 10^{-1}$   &  $6.4 \times 10^{-21}$   &  $1.3 \times 10^{38}$ & 48.6\\
                   &       &   $p$   &   5   &  $4.6 \times 10^{-1}$   &  $6.1 \times 10^{-21}$   &  $7.5 \times 10^{38}$ & 46.0\\
\hline
\hline
\end{tabular}
\end{table*}

Given the results obtained for  the quadrupole $g$-modes studied up to
now it is  natural to wonder whether other modes,  namely the $f$- and
$p$-modes,  of pulsating white  dwarfs can  radiate away  a measurable
amount of  gravitational waves.  To  this regard we have  extended our
calculations  to   incorporate  such   modes,  despite  the   lack  of
observational evidence for them. Obviously,  for these modes we do not
know which is the appropriate value of $\delta R_\star/R_\star$, since
the  estimate $\delta  R_\star/R_\star=10^{-4}$ is  based  on observed
$g$-modes in white dwarfs. However, for the calculations reported here
we adopted the same value. Fig.~2 shows the run of the functions $y_3$
and $y_4$ for  $f$-modes of our fiducial model,  for BPM~37093 and for
PG~1159-035.  As  is the  case for all  $f$-modes the radial  order is
zero, and their respective periods  are (from top to bottom): 11.35~s,
3.7~s and 20.9~s,  respectively. We note that now  the functions $y_3$
and   $y_4$   are   much   larger   than  in   the   case   previously
studied. Moreover, it is worth  noting that in this case the functions
$y_3$ and $y_4$ have large  amplitudes everywhere and do not vanish at
the  surface.   Consequently,  we   expect  that  a  large  amount  of
gravitational waves can be radiated  away. This is indeed the case, as
can be  observed in Table~2, where  we show the  model (first column),
its  respective  mass  (second  column), the  considered  mode  (third
column), the radial order (fourth column), the respective frequency of
the mode (fifth column),  the dimensionless strain (sixth column), the
luminosity radiated  away in the form of  gravitational waves (seventh
column) and  the kinetic energy (last  column). Note that  for all the
$f$-modes of the three models presented here the dimensionless strains
are several orders of magnitude larger and, moreover, the luminosities
radiated away are  much larger than the optical  luminosities, even of
the order of $10^{41}$~erg/s in the case of BPM~37093.  This, in turn,
could have  important consequences since  it could provide one  of the
possible reasons why  these modes have not been  observed thus far: if
they are excited they are  quickly damped by emission of gravitational
waves.

\begin{figure} 
\centering
\includegraphics[clip,width=250pt]{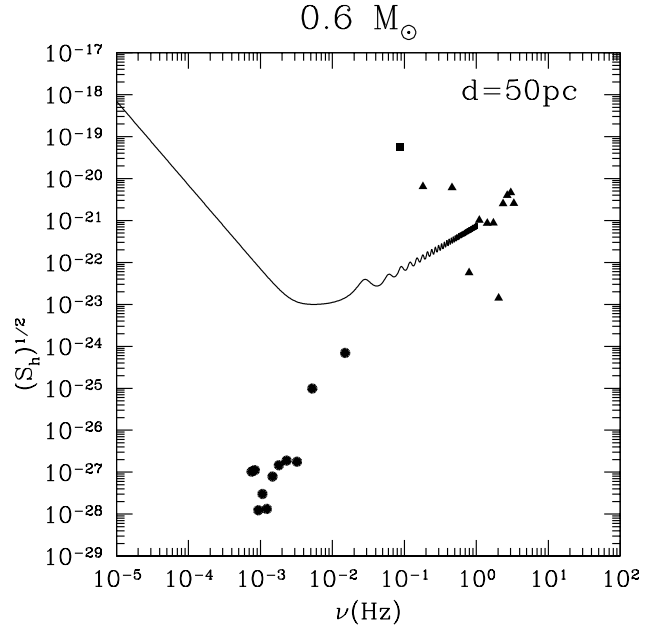}
\caption{A  comparison  of  the  signal  produced  by  the  quadrupole
$g$-modes of  our fiducial model ---  circles --- by  the $f$-mode ---
square ---  and by the $p$-modes  --- triangles ---  with the spectral
distribution of noise  of LISA for a one-year  integration period, and
assuming that the source is located at 50~pc.}
\label{fig4}
\end{figure}

Now we turn our attention to  the $p$-modes. Again, we show the run of
the $y_3$  and $y_4$ functions of  the $p$-modes in terms  of the mass
coordinate for our three models in  Fig.~3. We have chosen to show the
$k=25$  mode  for  all  three  cases.  Their  respective  periods  are
$P=0.56$~s  for our  fiducial $0.6\,  M_{\sun}$ model,  $P=0.11$~s for
BPM~37093 and $P=2.26$~s for PG~1159-035.  We note that the amplitudes
of the $y_3$  and $y_4$ functions for $p$-modes  are much smaller than
those  of   the  corresponding   $f$-modes,  and  comparable   to  the
corresponding $g$-modes studied before.  In addition, in contrast with
the situation for the $f$-modes, the amplitudes of $y_3$ and $y_4$ are
almost negligible in regions close  to the surface of the white dwarf.
However,   because  the   pulsation  frequencies   of   $p$-modes  are
considerably  higher  than  those  of  the  $f$-  and  $g$-modes,  the
dimensionless strains  --- see Table~2 --- are  consequently large and
the corresponding  gravitational wave  luminosities are very  large as
well,  although roughly  one  order of  magnitude  smaller than  those
obtained for the $f$-modes.

In order  to check  whether or not  LISA would  be able to  detect the
pulsating white dwarfs studied here  we have proceeded as follows.  We
first  have assumed  that the  integration time  of LISA  will  be one
year. The signal-to-noise ratio, $\eta$, is given by:

\begin{equation}
\eta^2=\int_{-\infty}^{+\infty}\frac{\tilde{h}^2(\sigma)}{S(\sigma)}
\frac{d\sigma}{2\pi}
\end{equation}

\begin{figure} 
\centering
\includegraphics[clip,width=250pt]{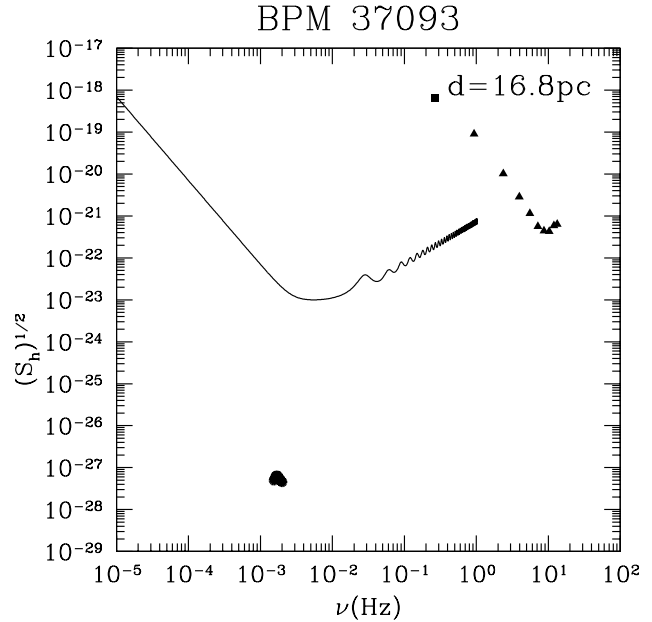}
\caption{Same as  Fig.~4 for  the case of  BPM~37093. The  distance in
this case is known, $d=16.8$~pc.}
\label{fig5}
\end{figure}

\begin{figure} 
\centering
\includegraphics[clip,width=250pt]{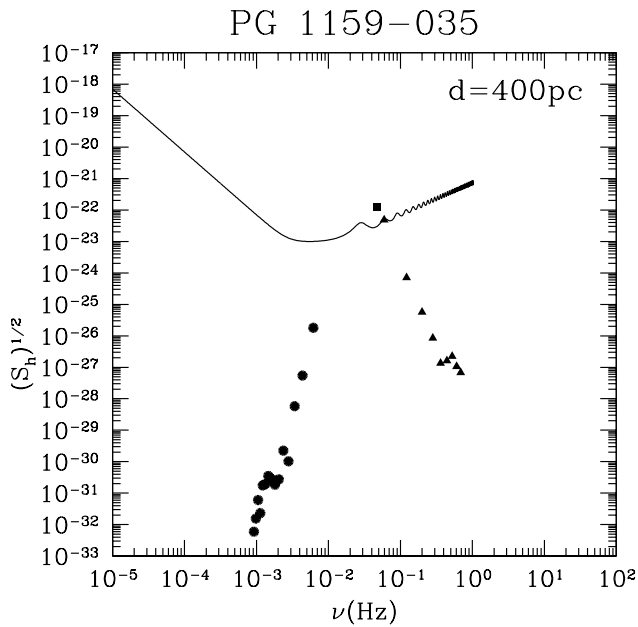}
\caption{Same  as  Fig.~4  for  the  case of  PG~1159-035.  Since  the
distance  to  PG~1159-035 is  not  accurately  known  we have  adopted
$d=400$~pc, which is a reasonable estimate.}
\label{fig6}
\end{figure}

\noindent where  $S(\sigma)=S_{\rm h}(\sigma)\tau$ is  the sensitivity
of LISA, $\tau$ is  the integration period, $\tilde{h}(\sigma)$ is the
Fourier Transform  of the dimensionless strain, and  $\sigma$ has been
previously defined.  It  can easily be shown that  for a monochromatic
gravitational wave $\eta=h(\sigma)/{S_{\rm h}^{1/2}(\sigma)}$. We have
adopted a  signal-to-noise ratio  $\eta=5$.  We have  furthermore used
the   integrated   sensitivity  of   LISA   as   obtained  from   {\tt
http://www.srl.caltech.edu/$\sim$shane/sensitivity}.   The  results of
this   procedure  are  shown   in  Figs.~\ref{fig4},   \ref{fig5}  and
\ref{fig6}  for  our fiducial  model  ($0.6\, M_{\sun}$  carbon-oxygen
white dwarf), for BPM~37093 and for PG~1159-035, respectively.  In all
three  figures the  $g$-modes are  shown as  circles, the  $f$-mode is
shown as  a square and the  $p$-modes are displayed  as triangles.  As
previously done, for our fiducial  model we have adopted a distance of
50~pc, for PG~1159-035  we have assumed a distance  of 400~pc, and for
BPM~37093 we  have used  its measured distance  (16.8~pc).  As  can be
seen, for  none of the three cases  studied here LISA will  be able to
measure the dimensionless strains of  the $g$-modes, even at a reduced
signal-to-noise ratio.  Most of the  $p$-modes will not be observed as
well, either because their frequencies  are too high to be observed by
LISA or  because they are too  weak. Moreover, given  that these modes
radiate huge amounts of energy in the form of gravitational waves they
become quickly damped and, consequently,  it will be very difficult to
detect them.  Particularly, and given  that for the  $p$-modes studied
here  the  time-averaged  dissipation   rates  of  pulsations  due  to
radiative (photons) heat leakage  and neutrino losses are much smaller
than the  luminosity radiated as  gravitational waves, an  estimate of
the  damping timescale,  $\tau_{\rm  d}$, can  be  easily obtained  by
considering the kinetic energy of the mode, $E_K$, which can be easily
computed from our numerical models:

\begin{equation}
\tau_{\rm d}\simeq\frac{2 E_K}{L_{\rm GW}}
\end{equation}

\noindent For instance,  for the $p$-mode with $k=1$  of BPM~37093 one
obtains  $\tau_{\rm d}\simeq 0.5$~yr,  which clearly  is too  short to
allow a detection.

Finally, the three  $f$-modes of the models presented  here lay in the
appropriate range of frequencies and, additionally, they are well over
the sensitivity  curve of LISA.  However,  as it was the  case for the
$p$-modes, they also radiate very large amounts of gravitational waves
and, hence, they will be  quickly damped, hampering the possibility of
detection.  In this  case  we obtain  a  damping timescale  $\tau_{\rm
d}\simeq 2.9$~yr for the $f$-mode of BPM~37093. Note that although the
luminosity radiated away in the  form of gravitational waves is larger
for  the $f$-mode  than  for the  $p$-mode  considered previously  the
damping timescale is  larger. This is so because  the kinetic energies
involved  are quite  different: $E_K=1.54\times  10^{49}$~erg  for the
$f$-mode and $E_K=6.42\times 10^{47}$~erg for the $p$-mode with $k=1$.
For the  sake of completeness we  present in Table~2  all the relevant
information  for the  $p$-  and $f$-modes  which  could be  eventually
detected.

\section{Discussion and conclusions}

In  this paper  we have  computed the  gravitational wave  emission of
pulsating white dwarfs. We have started by computing the gravitational
wave   radiation  of  white   dwarfs  undergoing   nonradial  $g$-mode
pulsations,  which are currently  observed in  a handful  of pulsating
white dwarfs. We have focused on three model stars. Our fiducial model
corresponded  to an  otherwise  typical $0.6\,  M_{\sun}$ model  white
dwarf with a carbon-oxygen fluid core and a hydrogen envelope. We have
also   paid  attention   to   two  additional   white  dwarf   models,
corresponding to  two stars for  which quadrupole $g$-modes  have been
observed  so far, namely,  BPM~37093 and  PG~1159-035.  We  have shown
that in all  these cases the gravitational wave signal  is too weak to
be observed  by future  space-borne interferometers, like  LISA.  More
specifically,  we have  found that  the  luminosities in  the form  of
gravitational waves radiated away by these stars and the corresponding
dimensionless strains  are very small  in all the cases,  in agreement
with the pioneering work of  Osaki \& Hansen (1973).  Hence, all these
sources contribute  to the Galactic noise and  no individual detection
are expected,  despite the proximity of  the sources. For  the sake of
completeness  we have  also elaborated  on  this subject  and we  have
computed the  gravitational wave  emission of white  dwarfs undergoing
nonradial $f$- and $p$-mode oscillations, even if these modes have not
been  observationally detected  whatsoever.   We have  found that  for
white dwarfs  undergoing these kind of pulsations  the luminosities in
the form  of gravitational waves radiated  away are very  large in all
the cases, in line with the earlier results of Osaki \& Hansen (1973).
Consequently,  these modes,  if excited,  should be  very short-lived,
thus hampering their eventual detection.

It may  seem that  for the case  of pulsating white  dwarfs undergoing
$g$-mode  oscillations there  could  still be  a  possibility of  {\sl
indirect detection}  by measuring  the secular rate  of change  of the
period   of   the  observed   modes.   However,   this   is  not   the
case. Particularly,  it can be easily  shown that the  secular rate of
change of the period of a pulsating white dwarf is given by

\begin{equation}
\frac{\dot P}{P}=-a \frac{\dot T}{T}+b \frac{\dot R_\star}{R_\star}
\label{pdot}
\end{equation}

\noindent where $T$ is the  temperature of the isothermal core and $a$
and  $b$ are constants  of order  unity which  depend on  the chemical
composition, thicknesses  of the H atmosphere and  He buffer, equation
of  state, and  other ingredients  involved in  the modeling  of white
dwarfs.  For  DA white  dwarfs in the  ZZ~Ceti instability  strip, the
second  term of  the right  hand side  of Eq.~(\ref{pdot})  is usually
negligible and,  thus, the secular rate  of change of  the period only
reflects  the speed of  cooling ---  see, for  instance, Isern  et al.
(1992) and references  therein --- whereas for GW~Vir  stars this term
is relevant, but  can be accounted for by  the theoretical models and,
hence,  the  speed  of  cooling  can also  be  derived.  However,  any
additional source  of extra cooling  --- like gravitational  waves ---
would  eventually translate into  an anomalous  rate of  period change
(Isern et al. 1992):

\begin{equation}
\frac{\dot P_{\rm o}}{\dot P_{\rm c}}-1= \frac{L_{\rm GW}}{L+L_\nu}
\label{c-o}
\end{equation}

\noindent where $P_{\rm o}$ is the observed period, $P_{\rm c}$ is the
computed  period   without  taking   into  account  the   emission  of
gravitational waves,  $L_\nu$ is the neutrino luminosity  --- which is
important for  hot white dwarfs ---  and the rest of  the symbols have
been previously  defined. The rate  of secular period change  has been
measured for some pulsating white dwarfs.  Particularly, for G117-B15A
(Kepler  et al.  2000)  it has  been possible  to measure  the secular
variation of the  main observed period of 215.2~s, $\dot  P = (2.3 \pm
1.4  ) \times 10^{-15}$~s~s$^{-1}$,  with unprecedented  accuracy.  In
fact, this white dwarf is the most stable optical clock known, and has
been  used  to  pose  tight  constraints  on the  mass  of  the  axion
(C\'orsico et al.   2001b) and the rate of  variation of gravitational
constant (Benvenuto  et al. 2004).   Other pulsating white  dwarfs ---
like L~19-2 and R~548 --- have also determinations of the secular rate
of period change but are not  as accurate as that of G117-B15A.  Note,
however, that for a $0.6\, M_{\sun}$ white dwarf undergoing quadrupole
$g$-mode oscillations  with a period  of $P\sim 200$~s the  right hand
side  of Eq.~(\ref{c-o}) amounts  to $2  \times 10^{-10}$  and $P_{\rm
c}\sim  10^{-15}$~s~s$^{-1}$.    In  other  words   the  gravitational
radiation will  produce a change in  $\dot P\sim 10^{-25}$~s~s$^{-1}$,
making  impossible  such  an   indirect  detection  even  if  accurate
observational data  and reliable theoretical  models become eventually
available. Again, the only case of  interest here would be the case in
which $L_{\rm GW}\sim  L + L_\nu$, which may be true  for the $f$- and
$p$-mode pulsators.

\begin{acknowledgements}
Part of  this work was  supported by the MCYT  grants AYA04094--C03-01
and 02,  by the European  Union FEDER funds,  and by the  CIRIT. L.G.A
acknowledges the  Spanish MCYT for  a Ram\'on y Cajal  Fellowship.  We
thank S.O.  Kepler  for his kind help in  some observational issues of
pulsating  white   dwarfs.   We  also  thank  J.A.   Pons  for  useful
discussions about gravitational waves and to our referee M. Montgomery
for constructive criticisms.
\end{acknowledgements}

\end{document}